\documentstyle[12pt]{article}
\newfont{\titlefont}{lcmss8 scaled 2986}
\renewcommand{\baselinestretch}{1}    
\def\e{\varepsilon}
\def\ec{\varepsilon_c}
\def\Tc{{\rm T}_c}
\def\bc{\beta_c}
\def\bm{\beta (\varepsilon)}
\def\a{\alpha}

\def\Tm{{\rm T}}
\def\tm{{\rm t}}
\def\t{{\rm t}}
\def\T{{\rm T}}

\def\be{\begin{equation}}
\def\ee{\end{equation}}
\begin{document}


\thispagestyle{empty}
\vspace*{3cm}
\begin{center}
\titlefont
On a trivial aspect of \\ [1.5ex]
canonical specific heat scaling
\vfill
\vfill
\vfill
\large
Michael Promberger\\ [1.5ex]
\large
Institut f\"ur Theoretische Physik, Universit\"at Erlangen-N\"urnberg,
Staudtstrasse 7, D--91058 Erlangen, Germany

\end{center}
\rm\normalsize

\vspace{2cm}
\begin{center}
\begin{minipage}{14cm}
{\bf Abstract.} We show that the canonical finite size
scaling of the specific heat emerges naturally -- and in some sense
trivially -- from the assumption that the microcanonical specific entropy
exhibits no substantial system size dependence.

\end{minipage}
\end{center}

\newpage
\section{Introduction}
%
Since the introduction of the Renormalization--Group by Wilson
\cite{Wilson}, we have learned much about critical phenomena and the striking
feature of universality. 
Unfortunately, the majority of systems
with nontrivial behaviour cannot be treated analytically.
Nevertheless, there exist approximation schemes concerning 
finite systems which allow to estimate
the properties of the corresponding infinite system by
proper extrapolation to the thermodynamic limit. A very powerful method
for the investigation of the critical properties of an infinite system
along these lines is the socalled finite size scaling theory introduced
by Fisher et al.~ (see \cite{Fisher}, \cite{Barber} and references therein).
Although hypothesised before the advent of the Renormalization--Group,
finite size scaling may be understood within the framework of the latter.\\
The main result of finite size scaling theory may be stated like this:
In the vicinity of the critical point of a given
infinite system, the system size dependence of certain
thermal properties of the corresponding {\em finite} system is governed by
properties (namely: the critical indices) of the {\em infinite} system.
Or, formulated slightly differently: In spite of the fact, that the free
energy density of a finite system is a completely analytic function of
its variables, the system size dependence of certain derivatives of the
free energy density are dictated by quantities which describe the
non--analytic behaviour of the free energy density of the corresponding
infinite system.

All thermal properties of a finite system (volume%
\footnote{Throughout this paper, we will use the language of lattice
systems with discrete energies, where the number of degrees of
freedom $N$ is equivalent to the volume $V$.}%
 $\,\,V\equiv N:=L^{\rm d}$) are
given by logarithmic derivatives of the canonical partition function
\be
Z_N(\T)\,:=\,\sum \limits_{x\in\Gamma}\,e^{-\beta H(x)} \qquad ,
\quad \T=\frac{1}{\beta} \qquad ,
\quad {\rm k}_B \equiv 1 \qquad ,
\ee
where $H(x)$ represents the energy of a particular microstate $x$ of the
system and the sum runs over all possible microstates which constitute
the space $\Gamma$ of all the states available to the system.
With the definition of the microcanonical density of states
\be
  \Omega_N (\e) \,:=\,\sum \limits_{x\in\Gamma}\,\delta_{H(x),N\e}
\ee
and the microcanonical specific entropy
\be
  \hat s_N(\e)\,:=\, \frac{1}{N}\,\ln \Omega_N(\e) \qquad ,
\ee
the canonical partition function reads
\be \label{partition_function}
Z_N(\T)\,:=\,\sum\limits_{\e}\,e^{N\left( \hat s_N(\e)-\beta \e \right)} \qquad
.
\ee
Here, the sum runs over all possible energy--values of the finite system.
As it is clearly visible, the system size dependence of the partition
function is due to two causes. Namely, the size dependence of the
microcanonical specific entropy $\hat s_N (\e)$ and the overall factor
$N$ in the exponential, which we will call the trivial system size
dependence.

It is the aim of this paper to demonstrate that finite size scaling emerges
naturally and in some sense trivially from the assumption that the critical
properties of the infinite system are already contained in the microcanonical
specific entropy of the finite system. Unfortunately, up to now we are
able to show this only as far as the finite size scaling properties of
the specific heat are concerned.

Although we are never introducing a specific system explicitly by giving
its Hamiltonian, we restrict our discussion to systems with short range
interactions. For this reason, standard finite size scaling is
applicable and hyperscaling holds%
\footnote{At least if the dimensionality of the system is smaller than
the upper critical dimension of the corresponding universality--class.}.

Remark:
In the thermodynamic limit, the entropy $\hat s_N(\e)$ is replaced by
the Massieu--function $\hat s(\e,h/\T)$ at zero magnetic
field $h\,\,$ $(=:\hat s(\e))$ , which is the Legendre transform of the
entropy $s(\e,m)$:
\be
\lim_{N \to \infty} \hat s_N(\e) \,\,\,=\,\,\,
\hat s(\e, \frac{h}{\T}=0) \,\,\,:=\,\,\,
\left. \sup \limits_m \left\{ s(\e,m) + \frac{h}{\T} \, m \right\}
\right|_{\frac{h}{\T}=0}
\ee
Here, $m$ denotes the magnetization per particle.
In this paper, the thus defined Massieu--function will be called
"microcanonical specific entropy".

In section \ref{s_infty}, we give an explicit form for the
microcanonical specific entropy of a system which undergoes a continuous
phase transition with a power law singularity of the specific heat.
In section \ref{s_scale}, we show that the system size independence
of the microcanonical specific entropy implies canonical finite size 
scaling. Namely, the scaling of the specific
heat maximum (section \ref{c_max_scale}) and the scaling of the
softening of the specific heat singularity (section
\ref{c_broad_scale}).
Since we are not able to proof the reverse direction of this
statement, in section \ref{hints}, we give some hints onto the
validity of the conjecture that the system size dependence of the
microcanonical specific entropy is such weak not to impact the canonical
finite size scaling.
%
%
\section{Microcanonical specific entropy $vs.$~continuous phase
transitions \label{s_infty}}
The microcanonical specific entropy of a system which undergoes a
continuous phase transition
may be written as a sum of a singular part $\hat s_{sing}(\e)$
and a correction term $\hat s_{corr}(\e)$ which is needed to correctly
describe the behaviour of
the specific entropy outside the critical region.
\be \label{s_sing_reg}
\hat s(\e) \, = \, \hat s_{sing}(\e) \, + \, \hat s_{corr}(\e)
\ee
If the corresponding specific heat shows a power law singularity, the choice
\be \label{s_sing}
\hat s_{sing}(\e)\,=\,\hat s_c+\bc(\e-\ec)-\frac{\bc |\ec|}{g}
\left|
\frac{\e-\ec}{\ec}
\right|^g
\left(\vphantom{\int}
\Theta(\ec-\e)A+\Theta(\e-\ec)A^\prime
\right)
\ee
(with $g:=\frac{2-\a}{1-\a}$)
for the singular part of the specific entropy yields the correct
behaviour of the singular part of the specific heat%
\footnote{
In the case of a logarithmic singularity the last term in $\hat s_{sing}(\e)$
should be replaced by a function of the form
$
\,\,\left(\e-\ec\right)^2\,/\,\ln \left|\e-\ec\right|\,
$.}.
$\hat s_c$, $\ec$ and
$\bc$ are the values of the specific entropy, the specific energy and
the inverse temperature at the critical point ($\bc:=1/\Tc$),
the step--function $\Theta(x)$ is defined by
$\Theta(x):=1$ $\forall \,x>0$ and $\Theta(x):=0$ $\forall \, x\le0$.
Indeed, since the microcanonical specific heat is given by
\be
c(\e):=-\frac{\bm^2}{s^{\prime\prime}(\e)}:=
-\frac{s^\prime(\e)^2}{s^{\prime\prime}(\e)}\quad ,
\ee
differentiating the singular part (\ref{s_sing}) of the specific entropy
twice with respect to the specific energy $\e$ yields
\be
c_{sing}(\e)\,=\,\frac{\bc |\ec|}{g-1}\,
\left(
\frac{\Theta(\ec-\e)}{A}+\frac{\Theta(\e-\ec)}{A^\prime}
\right)\,\,
\left|
\frac{\e-\ec}{\ec}
\right|^{-\frac{\a}{1-\a}}
\ee
in the vicinity of the critical point $\ec$.
Alternatively, by going over from $\left|\frac{\e-\ec}{\ec}\right|$
to $\left|\frac{\Tm-\Tc}{\Tc}\right|$ via
\be
\left|
\frac{\bm-\bc}{\bc}
\right|\,
\approx \,
\left|\frac{\e-\ec}{\ec}\right|^\frac{1}{1-\a}
\left( \vphantom{\int} \Theta (\ec-\e) A + \Theta (\e-\ec ) A^\prime \right) \quad ,
\ee
the singular part of the microcanonical specific heat as a function of
the reduced temperature $\tm:=(\Tm-\Tc)/\Tc$
is proportional to $|\tm|^{-\a}$:
\be \label{c_sing}
c_{sing}(\tm)=(1-\a)\frac{|\ec|}{\Tc}
\left(
\frac{\Theta(\Tc-\Tm)}{A^{1-\a}}+\frac{\Theta(\Tm-\Tc)}{{A^\prime}^{1-\a}}
\right)\,
\left|\tm\right|^{-\a}
\ee
The correction term $\hat s_{corr}(\e)$ of the specific entropy yields no
contribution to the singular behaviour of the specific heat if it obeys
the following condition:
\be \label{s_reg_cond}
\lim_{\e\to\ec} \frac{\hat s_{corr}(\e)}{|\e-\ec|^g}\,=\,0
\ee
%
%
\section{Microcanonical specific entropy $vs.$~finite size scaling of the
canonical specific heat \label{s_scale}}

In the rest of this paper, we will study finite size scaling properties
of the canonical specific heat of a hypothetical $N$--particle system
with specific entropy
\be \label{postulat}
\hat s_N(\e) \, \equiv \, \hat s_{sing}(\e)\,+\,\hat s_{corr,N}(\e)
\qquad \forall \quad N\,>\,N_0 \qquad .
\ee
This implies the assumption that, at least for sufficiently large $N$,
the singular contribution to $\hat s_N(\e)$ is identical to the singular part
of the entropy of the infinite system.
The correction term $\hat s_{corr,N}(\e)$ may show some $N$--dependence
which should of course be consistent with the condition (\ref{s_reg_cond}).


Having postulated the form of the entropy in (\ref{postulat}) the
canonical specific heat $c_N(\T)$ of the $N$--particle system follows
directly from (\ref{partition_function}). We shall compare the scaling
properties of the thus determined specific heat $c_N(\T)$ with the
results of conventional finite size scaling theory \cite{Binder}.
At the critical temperature $\Tc$ of the infinite system the finite 
size scaling theory
predicts for the value of the specific heat of the finite
system
\be \label{hyper_k}
c_N(\Tc) \, \propto \, L^\frac{\a}{\nu} \qquad .
\ee
In the finite system the singularity is smeared. Scaling theory predicts
for the width of the specific heat anomaly:
\be \label{softening}
\Delta \T(L) \, \propto \, \left(\frac{1}{L}\right)^\frac{1}{\nu}
\qquad .
\ee
Here, $\nu$ is the critical exponent of the correlation length
$\xi (\t)$ of the infinite system, $L$ is
the linear dimension of the finite system and $\Tc$ denotes the critical
temperature of the infinite system.

We are now going to show, that the finite size scaling relations
\be \label{hyper_m}
c_N (\Tc) \, \propto \, L^{\frac{{\rm d}\a}{2-\a}}
\ee
and
\be \label{soft_m}
\Delta \T(L) \, \propto \,
\left(\frac{1}{L}\right)^\frac{\rm d}{2-\a}
\ee
are direct consequences of the postulate (\ref{postulat}). Together with
the validity of hyperscaling, (\ref{hyper_m}) implies (\ref{hyper_k}) and
(\ref{soft_m}) implies (\ref{softening}).
While (\ref{soft_m}) is established by numerical integration,
(\ref{hyper_m}) can be shown analyticaly.

\subsection{Canonical specific heat scaling at $\Tc$ \label{c_max_scale}}

The first step consists in the calculation of the
$n$--th moment of the specific energy of the $N$--particle
system with respect to the canonical distribution at the critical
temperature of the infinite system.
\be \label{n_moment}
 \left\langle
   \e^n
 \right\rangle_N (\Tc)\,\,=\,\,
 \frac{1}{Z_N(\Tc)} \sum_\e \,
 \e^n\,e^{N(\hat s_N(\e)-\bc \e)}
\ee
$Z_N(\Tc)$ denotes the canonical partition function of the
$N$--particle system at $\Tc$ (cf.~(\ref{partition_function})).
For sufficiently large $N$, it is justified to replace the sum over all
possible energy--values by an integration along the
energy--axis. Since the correction term $\hat s_{corr,N}(\e)$ of the
specific entropy will
yield no contribution to the canonical quantities at $\Tc$
(for sufficiently large $N$ again), we are concerned with integrals of
the type
\be
{\rm I}_n
\, := \,
\int\limits_{-\infty}^\infty d\e \,
\e^n \,
\exp \left\{ -N\,\frac{\bc |\ec |}{g}\, \left| \frac{\e-\ec}{\ec} \right|^g
\,\left(\vphantom{\int}\Theta (\ec - \e)A + \Theta (\e-\ec) A^\prime \right)\, \right\}
\ee
Defining new amplitudes $B:=A\bc|\ec |^{1-g}/g$,
$\,\,B^\prime:=A^\prime \bc|\ec |^{1-g}/g$ and renaming
$(\e-\ec) \rightarrow \e$, we get
\be
{\rm I}_n\,=\,
\sum_{l=0}^n \, {n \choose l} \, \ec^{n-l}\,
\int\limits_0^\infty d\e \, \e^l \,
\left(
  e^{-NB^\prime \e^g} + (-1)^l e^{-NB\e^g}
\right)
\ee
Substituting $x:=NB^\prime \e^g$ in the first and $x:=NB\e^g$ in the
second term of the integral,
we end up with the following expression for ${\rm I}_n$:
\begin{eqnarray}\nonumber
{\rm I}_n & = &
\sum_{l=0}^n {n \choose l} \, \frac{\ec^{n-l}}{g}
\left(
 {B^\prime}^{-\frac{l+1}{g}} + (-1)^l B^{-\frac{l+1}{g}}
\right)
N^{-\frac{l+1}{g}}
\int\limits_0^\infty dx \, x^{\frac{l+1}{g}-1}\,e^{-x} \\ [1ex]
& = &
\sum_{l=0}^n {n \choose l} \, \frac{\ec^{n-l} \Gamma \left(
\frac{l+1}{g} \right)}{g}
\left(
 {B^\prime}^{-\frac{l+1}{g}} + (-1)^l B^{-\frac{l+1}{g}}
\right)
N^{-\frac{l+1}{g}}
\end{eqnarray}
Since we want to study the finite size scaling properties of the
canonical specific heat $c_N(\T)$
\be \label{c_kan}
c_N(\Tc) \, := \,
N \bc^2
\left(
\left\langle \e^2 \right\rangle_N(\Tc)\,-\,
\left\langle \e^{\vphantom{2}} \right\rangle^2_N (\Tc)
\right)
\ee
at the critical temperature $\Tc$ of the infinite system, we have to look
at the second central moment of the specific energy with respect to the
canonical distribution:
\begin{eqnarray} \label{cent_mom}
\lefteqn{ \hspace{-3cm}
\left\langle \e^2 \right\rangle_N(\Tc)\,-\,
\left\langle \e^{\vphantom{2}} \right\rangle^2_N(\Tc)
\, = \,\frac{\rm I_2}{\rm I_0}\,-\,\left( \frac{\rm I_1}{\rm I_0}
\right)^2 \,\, =
} \nonumber \\[2ex]
\phantom{m} \hspace{2cm} & = &
N^{-\frac{2}{g}}\,\,
\left[
\frac{\Gamma (\frac{3}{g})}{\Gamma (\frac{1}{g})}
\,\,
\frac{{B^\prime}^{-\frac{3}{g}}+B^{-\frac{3}{g}}}{{B^\prime}^{-\frac{1}{g}}+B^{-\frac{1}{g}}}
\,-\,
\left(
\frac{\Gamma (\frac{2}{g})}{\Gamma (\frac{1}{g})}
\,\,
\frac{{B^\prime}^{-\frac{2}{g}}-B^{-\frac{2}{g}}}{{B^\prime}^{-\frac{1}{g}}+B^{-\frac{1}{g}}}
\right)^2
\right]\,\propto\,N^{-\frac{2}{g}} \phantom{space}
\end{eqnarray}
With $g:=(2-\a)/(1-\a)$ and $N=L^{\rm d}$, {\rm d} being the dimension
of configuration space, eq.~(\ref{hyper_m}) follows immediately from
(\ref{cent_mom}).
In conjunction with the hyperscaling relation
\be \label{hyper}
{\rm d}\nu \, = \, 2-\a \qquad ,
\ee
this implies (\ref{hyper_k}).

\subsection{Scaling of the specific heat width $\Delta\T$
\label{c_broad_scale}}
The specific heat singularity (\ref{c_sing}) of the infinite system is
rounded
in the corresponding finite systems.
Since in the canonical ensemble there are no phase transitions in
finite systems, this effect seems to be quite natural. The standard
(finite size scaling) argument for this softening is that the specific
heat of the finite system saturates for those temperatures, where
the correlation length $\xi(\t)$ of the infinite system becomes larger than
the linear system size $L$. The correlation length $\xi_N(\t)$ of the finite
system is bounded from above by a length which is of the order of
magnitude of the linear system size $L$.
For this reason, there is a temperature region within which the specific
heat of the finite system deviates essentially from the specific heat of
the infinite system.\\
Any measure of the width of this region will do.
For numerical convenience, the width $\Delta  \T (L)$ is defined as the
temperature range were $c_N(T)$ is larger than 80\% of its maximum
value.
Having defined $\Delta \T(L)$, it is an easy matter to compute it
by numerical integration via eqs.~(\ref{partition_function}) and
(\ref{s_sing}) for any lattice size $L$.
The various parameters appearing in (\ref{s_sing}) have
not been chosen arbitrarily but we have taken the parameter set obtained
by a fit to the (simulated) entropy data of a three--dimensional Ising
model with linear system size $L=18$. The parameters are%
\footnote{The value of $\hat s_c$ is not listed, because an additive
          constant in the specific entropy is quite irrelevant with
          respect to the physics described by that entropy. Likewise the
          value of $\ec$ is irrelevant for the smearing.}%
: $\ec=-1.059$; $\a=0.1155$; $\beta_c=0.222684$; $A=0.091$;
$A^\prime=0.150$.\\
Fig.~1 shows a log--log plot of $\Delta \T (L)$ {\em vs.~}$1/L$ where we
have chosen $N=L^3=10^8,...,10^{10}$ together with a straight--line fit
to the data points.
The critical exponent $(2-\a)/{\rm d}$, which is just the inverse
slope of the fitted straight line (cf.~(\ref{soft_m})), emerges to be
$(2-\a)/{\rm d}=0.6275$ which is consistent with the value of $\nu=0.6282$
obtained by combining the input--value of $\a=0.1155$ with the
hyperscaling relation (\ref{hyper}).
\section{\label{hints}On the system size dependence of microcanonical specific entropies}
In the previous section, we have shown that a system size independent microcanonical
specific entropy implies the canonical finite size scaling relations. Unfortunately,
we are not able to proof this statement in the reverse direction.
Nevertheless, we can report at least two observations which are necessary
(not sufficient) for the validity of the statement that the system size
dependence of the microcanonical specific entropy has no considerable 
impact on canonical finite size scaling (if the systems are not chosen
to be too small).

1) If the microcanonical specific entropy shows no system size dependence and
if the critical properties of the infinite system are already contained in the
entropy of the corresponding finite systems, then it should be possible to
extract information about the critical exponent $\alpha$ by performing some fits
of a function of the form (\ref{s_sing}) to the entropy data of finite systems
obtained by, e.g., Monte Carlo simulation.\\
And indeed, it has already been shown \cite{Promberger},
that an entropy of the form (\ref{s_sing_reg}) with the singular part
given in (\ref{s_sing}) fits
the data of a $10 \times 10 \times 10$--Ising system very well and
it will be shown elsewhere, that the same entropy with the same
exponent $g$ but slightly modified parameters $\ec$, $\bc$ is well
suited to fit the data of larger 3D--Ising systems
(the values of the critical exponent $\alpha$ emerging from this
fits is well consistent with the respective value of $\alpha$
in the thermodynamic limit, i.e.~$\alpha_{fit} \in [.08;.12]$).

2) If the microcanonical specific entropy shows no substantial system size 
dependence, then it should make no difference if the value of the specific
heat of a finite system at the critical temperature of the infinit system
is calculated by use of the entropy of the finite
system or by use of the entropy of the infinit system.\\
Fortunately, this can be checked in the case of the twodimensional Ising
model, where the entropy of the infinit system can be calculated from
Onsagers solution \cite{Onsager}. At zero external field, the internal energy per particle
as a function of the inverse temperature reads as:
\be \label{Onsager}
\epsilon(\beta)=-\coth (2\beta)\,
\left[
1+\frac{2}{\pi}
\left(
2\tanh^2 (2\beta)-1
\right)
K_1(q)
\right]
\ee
where
\be
q:=\frac{2\sinh (2\beta)}{\cosh^2(2\beta)}
\qquad
\mbox{and}
\quad
K_1(q):= \int \limits_0^{\pi /2}
{\rm d}\varphi
\left(
1-q^2 \sin^2\varphi
\right)^{1/2}  
\qquad , \quad J\equiv k_B \equiv 1 \quad .
 \ee
Here, $J$ denotes the Ising coupling constant.
Since the inverse temperature $\beta$ is defined to be the derivative
of the entropy $\hat{s}(\e)$ with respect to the energy $\e$, the entropy
can be calculated according to
\be
\hat{s}(\e) =
const. +
\int \limits_{\e_0}^\e {\rm d}\tilde{\e}\,\beta (\tilde{\e})
\ee
with arbitrary $\e_0$. $\beta (\e)$ is obtained by inverting eq.~(\ref{Onsager}).
In the case of a logarithmic specific heat singularity, the canonical finite size
scaling theory predicts \cite{Ferdinand}
\be \label{2dscale}
c_N(T_c) \propto \ln (1/L)
\ee
Having obtained the entropy of the infinit 2$D$ Ising system, it is an easy
thing to calculate the critical point specific heat using eq.~(\ref{c_kan}).
The result is shown in figure 2

\section{Conclusion}
We have shown that the finite size scaling relations (\ref{hyper_m})
and (\ref{soft_m})
are trivial consequences of the postulate (\ref{postulat}):
for sufficiently large $N$, the entropy of the finite system was
assumed to be identical to the entropy of the infinite system at least
in the vicinity of the critical point. In this context, "trivial"
means that the softening of the specific heat singularity
is caused solely by the trivial factor $N$ in the exponential
of the canonical partition function (\ref{partition_function}).
In the framework of this
scenario it is therefore not astonishing that some properties of the
finite system are governed by the critical indices of the infinite
system: they are already contained in the entropy of the finite
system but they are covered up by the averaging ("smearing") procedure
of the canonical partition function (for a detailed discussion of this
"smearing--effect" see \cite{Huller}).
For this reason it seems to be plausible that, as far as finite systems
are concerned, we are in some sense blinded by the canonical formalism
which obscures the information already available
in the microcanonical specific entropy.\\
Indeed, the hypothetical system which we have discussed may seem to be a
rather strange construction but it is not as arbitrary as it seems to be
since we have already shown that an entropy of the type (\ref{s_sing})
is well suited to fit the data of a $10^3$--3d--Ising system.
Note that this is by no means the only example of a system with a
microcanonical specific entropy $\hat s_N(\e)$ which shows $no$ substantial
$N$--dependence. We will report about other examples elsewhere.

\section{Acknowledgements}
The author wants to thank Alfred H\"uller and Hajo Leschke
for many stimulating discussions.

%

\newpage
\phantom{llllllllllll}
\hspace{3cm}

\centerline{\underline{Captions}}

\vspace{5cm}

{\bf Fig.~1} $\log$--$\log$ plot of $\Delta \T (L)$ {\em vs.~}$1/L$ for
system sizes $N=L^3=10^8,...,10^{10}$. A straight--line fit to the data
points yields $\,(2-\a)/{\rm d}=0.6275$ which is consistent with the
predicted value of $\nu=0.6282$.

\newpage
\phantom{llllllllllll}
\hspace{3cm}

\centerline{\underline{Figure 1}}

\vspace{5cm}

\begin{figure}[h]
\begin{center}
%
\setlength{\unitlength}{0.1bp}
\begin{picture}(3600,2160)(0,0)
\put(2008,-20){\makebox(0,0){$\log \frac{1}{L}$}}
\put(250,1180){%
\makebox(0,0)[b]{\shortstack{$\log \Delta {\rm T}$}}%
}
\put(3417,151){\makebox(0,0){-2.6}}
\put(3065,151){\makebox(0,0){-2.7}}
\put(2713,151){\makebox(0,0){-2.8}}
\put(2361,151){\makebox(0,0){-2.9}}
\put(2008,151){\makebox(0,0){-3}}
\put(1656,151){\makebox(0,0){-3.1}}
\put(1304,151){\makebox(0,0){-3.2}}
\put(952,151){\makebox(0,0){-3.3}}
\put(600,151){\makebox(0,0){-3.4}}
\put(540,2109){\makebox(0,0)[r]{-2.6}}
\put(540,1799){\makebox(0,0)[r]{-2.8}}
\put(540,1490){\makebox(0,0)[r]{-3}}
\put(540,1180){\makebox(0,0)[r]{-3.2}}
\put(540,870){\makebox(0,0)[r]{-3.4}}
\put(540,561){\makebox(0,0)[r]{-3.6}}
\put(540,251){\makebox(0,0)[r]{-3.8}}
\end{picture}
\end{center}
\end{figure}


\vspace{2cm}
\addcontentsline{toc}{section}{References}
\begin{thebibliography}{99}
  \bibitem{Wilson} K.~G.~Wilson and J.~Kogut; Phys.~Rep.~{\bf 12C}, 75
  (1974)
  \bibitem{Fisher} M.~E.~Fisher and M.~N.~Barber;
   Phys.~Rev.~Lett {\bf 28}, 1516 (1972)
  \bibitem{Barber} M.~N.~Barber in;
  "Phase Transitions and Critical Phenomena" Vol.~8 \\
  edited by Domb/Lebowitz, Academic Press (1983), pp 146
  \bibitem{Promberger} M.~Promberger and A.~H\"uller; Z.~Phys.~B {\bf
  97}, 341 (1995)
  \bibitem{Binder} K.~Binder; in
  "Finite Size Scaling and Numerical Simulation of Statistical Systems"\\
  edited by V.~Privman, World Scientific (1990), pp 173
  \bibitem{Onsager} L.~Onsager, Phys.~Rev.~{\bf 65}, 117 (1944)
  \bibitem{Ferdinand} A.E.~Ferdinand and M.E.~Fisher; Phys.~Rev.~{\bf 185}, 832 (1969)
  \bibitem{Huller} A.~H\"uller; Z.~Phys.~B {\bf 93}, 401 (1994)
\end{thebibliography}
\end{document}